\documentclass[aps,prl,preprint,preprintnumbers,amsmath,amssymb]{revtex4-1}
\usepackage{graphicx}
\usepackage{dcolumn}
\usepackage{bm}
\usepackage{amsmath}
\usepackage{color}

\begin{document}

\def\gsim{\mathop {\vtop {\ialign {##\crcr 
$\hfil \displaystyle {>}\hfil $\crcr \noalign {\kern1pt \nointerlineskip } 
$\,\sim$ \crcr \noalign {\kern1pt}}}}\limits}
\def\lsim{\mathop {\vtop {\ialign {##\crcr 
$\hfil \displaystyle {<}\hfil $\crcr \noalign {\kern1pt \nointerlineskip } 
$\,\,\sim$ \crcr \noalign {\kern1pt}}}}\limits}


\title{
%
Magnetism and topology in Tb-based icosahedral quasicrystal}


\author{Shinji Watanabe}
\affiliation{Department of Basic Sciences, Kyushu Institute of Technology, Kitakyushu, Fukuoka 804-8550, Japan 
}


\date{\today}

\begin{abstract} 
Quasicrystal (QC) possesses a unique lattice structure with rotational symmetry forbidden in conventional crystals. The electric property is far from complete understanding and it has been a long-standing issue whether the magnetic long-range order is realized in the QC. The main difficulty was lack of microscopic theory to analyze the effect of the crystalline electric field (CEF) at the rare-earth atom in QCs. Here we show the full microscopic analysis of the CEF in Tb-based QCs. We find that magnetic anisotropy arising from the CEF plays a key role in realizing unique magnetic textures on the icosahedron whose vertices Tb atoms are located at. 
Our analysis
of the minimal model based on the magnetic anisotropy 
suggests
that the long-range order of the hedgehog characterized by the topological charge of one is stabilized in the Tb-based QC.
We also find that the whirling-moment state is characterized by unusually large topological charge of three.  The magnetic textures as well as the topological states are shown to be switched by controlling compositions of the non-rare-earth elements in the ternary compounds. 
Our model is useful to understand the magnetism as well as the topological property in the rare-earth-based QCs and approximant crystals. 
\end{abstract}


\maketitle

Quasicrystal (QC), which was discovered in 1984~\cite{Shechtman}, has a unique lattice structure with the rotational symmetry forbidden in conventional crystals~\cite{Tsai,Takakura}. 
In the QC, it has been an open question whether the magnetic long range order is realized~\cite{Suzuki}. 
Thus far, in the rare-earth based approximant crystal (AC) retaining the periodicity as well as the local atomic configuration common to the QC, the magnetic long-range orders have been observed in Cd$_6$R (R=Nd, Sm, Gd, Tb, Dy, Ho, Er, and Tm)~\cite{Tamura2010,Mori,Tamura2012,Das} and Au-SM-R (SM=Si, Ge, Sn, and Al, and R=Gd, Tb, Dy, and Ho)~\cite{Hiroto2013,Hiroto2014}. 
Most of these measurements were performed by using the bulk probe such as susceptibility and the detailed magnetic structures remain unresolved.
Recently, in the Tb-based ACs, the magnetic structures have started to be identified by neutron measurements~\cite{Sato2019,Hiroto}. 
Quite recently, the ferromagnetic (FM) long-range order has been discovered experimentally in the Tb-based QC~\cite{Tamura2021}.

These experimental developments urgently require theoretical investigation of the magnetism in the rare-earth based QC and AC. 
Theoretical studies to date have been performed as the model calculations mostly by the spin model and also by the Hubbard model in a small cluster or low-dimensional systems~\cite{Okabe,Sorensen,Coffey,Jan,Axe,Wessel,Kons,Jan2007,Hucht,Thiem,Komura,Sugimoto,Koga2017,Koga2020,STS,Miyazaki}, where the effect of the crystalline electric field (CEF) crucial for the rare-earth system was not taken into account microscopically. 

%
In the rare-earth based compounds, the effect of the CEF is important for the electronic states, which has been well recognized in the periodic crystals such as the heavy fermion systems. In contrast, in the rare-earth based QC and AC, understanding about the CEF is very limited because the local symmetry around the rare-earth site is completely different from the crystallographic point group, e.g., 5-fold symmetry allowed in the QC and AC is forbidden in the periodic crystals. Thus far, the CEF in the rare-earth based ACs was analyzed experimentally~\cite{Hiroto,Jaz,Das}. Recently, the full CEF Hamiltonian in the rare-earth based QC and AC has been formulated theoretically~\cite{WM2021}.

%
In this article, by applying this formulation to the Tb-based QC and AC, we report full microscopic analysis of the CEF in the Tb-based QC and AC.
We show that the magnetic anisotropy arising from the CEF plays a key role in realizing unique magnetic textures on the icosahedron (IC) at whose vertices the Tb moments are located in the QC and AC. 
We propose the minimal model to clarify the magnetic structures in the QCs and ACs. 
Our analysis suggests
that the long-range order of the hedgehog characterized by the topological charge is realized in the Tb-based QC. This is the first theoretical discovery of the 
possible 
long-range order of the topological magnetic texture in the QC.
Distinct from the topological spin textures intensively studied so far in the periodic crystals~\cite{Taguchi,Nagaosa,Kanazawa,Fujishiro,Ishiwata,Tokura,Aoyama}, our discovery is based on the total angular momentum on the CEF-anisotropy-protected IC, which provides a new concept of {\it topological quasicrystal}.

\begin{figure}[t]
\includegraphics[width=8.9cm]{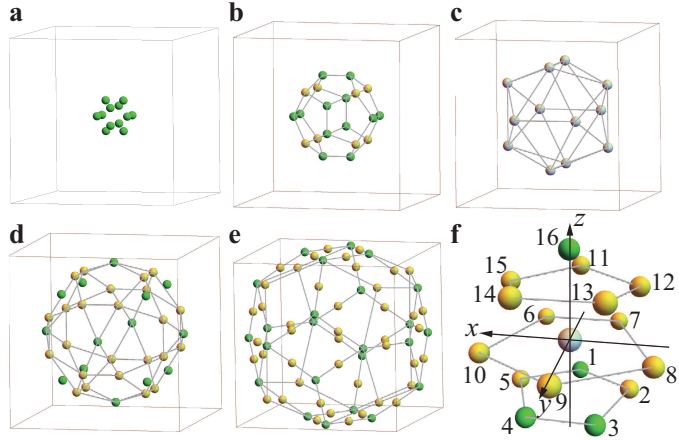}%
\caption{(color online) 
{\bf Crystal structure of Tsai-type cluster and local atomic configuration.} 
Tsai-type cluster consists of 
{\bf a}, the cluster center, 
{\bf b}, dodecahedron, 
{\bf c}, icosahedron, 
{\bf d}, icosidodecahedron, and 
{\bf e}, defect rhombic triacontahedron 
with Au (yellow), Au/Si (green), and Tb (gray). 
{\bf f}, Local atomic configuration around Tb. 
(This figure is created by using Adobe Illustrator CS5 Version 15.1.0.)
}
\label{fig:atoms}
\end{figure}

The Tb-based QC and AC consist of 
{building blocks named}
the Tsai-type cluster composed of the concentric shell structures shown in Figs.~\ref{fig:atoms}a-\ref{fig:atoms}e, where Au$_{70}$Si$_{17}$Tb$_{13}$ is displayed as a representative case. 
{Namely, a set of polyhedrons are nested like a matryoshka doll.}
In Fig.~\ref{fig:atoms}c, Tb is located at each vertex of the IC, which forms the Tb-12 cluster. 
The local atomic configuration around Tb is shown in Fig.~\ref{fig:atoms}f. 
Here, the $z$ axis is passing through Tb from the center of the IC, which is regarded as the pseudo 5-fold axis, and 
the $y$ axis is taken so as the $yz$ plane to be the mirror plane. 
{Namely, the configuration of atoms is symmetric in both sides with respect to the $yz$ plane in Fig.~\ref{fig:atoms}f.}

\noindent
{\bf Results}

\noindent
{\bf Analysis of crystalline electric field.} 
The CEF can be analyzed on the basis of the point charge model. 
Since the Tb$^{3+}$ ion has the $4f^{8}$ configuration, it is regarded to have the charge of 6$|e|$ in the hole picture ($4f^{14}$ is the closed shell of $4f$ electrons). 
Then, the electrostatic energy at the Tb$^{3+}$ ion is expressed as 
$H_{\rm CEF}=6|e|V$, where $V$ is the Coulomb potential $V=\sum_{i=1}^{16}q_i/|{\bm r}-{\bm R}_i|$. 
Here, ${\bm r}$ and ${\bm R}_i$ are the position vectors of the Tb$^{3+}$ ion and ligand ions respectively and $q_i$ is the charge at the $i=1$--16th Au or Au/Si mixed site shown in Fig.~\ref{fig:atoms}f. 
To take into account the effect of the Au/Si mixed sites, here we set $q_i=(0.63Z_{\rm Au}+0.37Z_{\rm Si})|e|$ for 
$i=1$, 3, and 4 and $q_{16}=(0.19Z_{\rm Au}+0.81Z_{\rm Si})|e|$ so as to reflect the occupancy ratio~\cite{Hiroto}, otherwise $q_i=Z_{\rm Au}|e|$, where $Z_{\rm Au}$ and $Z_{\rm Si}$ are the valences of Au and Si respectively. 

In metallic crystals and alloyed ACs, the valences of Au and Si are known to be 1 and 4 respectively~\cite{Pearson,Mizutani}. 
However, in reality, these values can be reduced by the screening effect of the conduction electrons. Hence, we analyze the CEF for $Z_{\rm Si}=\alpha Z_{\rm Au}$ by setting $Z_{\rm Au}=0.223$ as a typical value which was determined by the neutron measurement in Au$_{70}$Si$_{17}$Tb$_{13}$~\cite{Hiroto}. It is noted here that the choice of $Z_{\rm Au}$ does not affect the CEF eigenstate as long as $\alpha$ is the same.

\begin{figure}[t]
\includegraphics[width=8.9cm]{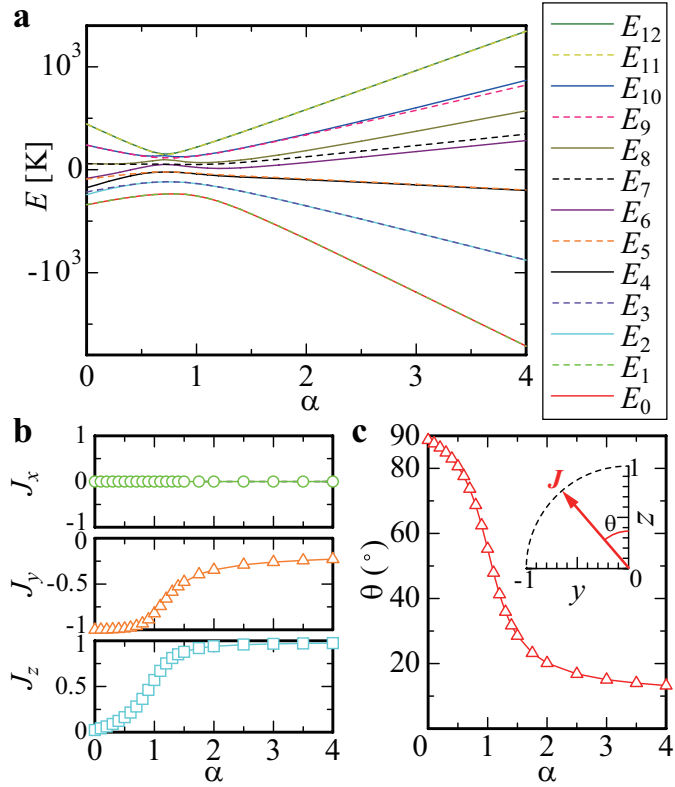}%
\caption{(color online) 
{\bf Crystalline-electric-field energies and magnetic-moment direction of the ground state.}  
{\bf a}, The $\alpha$ dependence of the CEF energies at Tb in $Z_{\rm Si}={\alpha}Z_{\rm Au}$. 
{\bf b}, The $\alpha$ dependence of the largest-moment direction of the CEF ground state shown by the normalized vector ${\bm J}$. (top panel) $J_x$ vs $\alpha$, (middle panel) $J_y$ vs $\alpha$, and (bottom panel) $J_z$ vs $\alpha$. 
{\bf c}, The $\alpha$ dependence of $\theta$ defined by the angle of ${\bm J}$ from the $z$ axis in Fig.~\ref{fig:atoms}f.
Inset illustrates the ${\bm J}$ vector in the $yz$ plane. 
(This figure is created by using Adobe Illustrator CS5 Version 15.1.0.)
}
\label{fig:CEF}
\end{figure}

On the basis of the recently developed formulation (Supplementary information)~\cite{WM2021}, 
we diagonalize $H_{\rm CEF}$ for the total angular momentum $J=6$ as the ground multiplet endorsed by the Hund's rule. 
Then, the CEF energies $E_n$ and the eigenstates $|\psi_n\rangle$ are obtained, as shown in Fig.~\ref{fig:CEF}a. 
The CEF energies are split into almost 7 for $\alpha=0$ to 8 for $\alpha=4$, some of which are nearly degenerate. 
The CEF ground energy $E_0$ is well separated from the first excited energy $E_1$ as seen in Fig.~\ref{fig:CEF}a, as indeed observed in Au$_{70}$Si$_{17}$Tb$_{13}$ with $E_1-E_0$ being in the order of $10^2$~K. 
Thus, the magnetism at low temperatures is dominated by the CEF ground state.

\noindent
{\bf Magnetic anisotropy.}  
Then we evaluate the largest magnetic-moment direction of the CEF ground state, i.e., the principal axis. 
The result, which is expressed as the normalized vector ${\bm J}=(J_x, J_y, J_z)$, is shown in Fig.~\ref{fig:CEF}b. 
This reveals that the magnetic moment is lying in the $yz$ plane i.e., the mirror plane in Fig.~\ref{fig:atoms}f and changes its direction depending on $\alpha$. 
Namely, as $\alpha$ increases from $0$, the magnetic moment rotates to anticlockwise direction in the mirror plane, whose angle defined from the $z$ axis, $\theta$, changes from $\theta\approx 90^{\circ}$ to approach $\theta=0^{\circ}$, as shown in Fig.~\ref{fig:CEF}c. 

\noindent
{\bf Minimal model on icosahedron.} 
This result indicates that the direction of the magnetic moment at each Tb site can be changed by controlling the compositions of Au and SM in the Au-SM-Tb ACs and QCs. 
To clarify how the magnetic anisotropy affects the magnetism in the IC, we consider the minimal model 
\begin{eqnarray}
H=-\sum_{\langle i,j\rangle}J_{ij}\hat{\bm J}_i\cdot\hat{\bm J}_j.
\label{eq:H}
\end{eqnarray}
Here, $\hat{\bm J}_i$ is the unit Ising-spin vector operator at the $i$th Tb site, whose direction is restricted to either parallel or antiparallel to the principal axis shown in Fig.~\ref{fig:CEF}b and $J_{ij}$ is the exchange interaction. 
We consider the nearest neighbor (N.N.) interaction $J_1$ and the next. N.N. (N.N.N.) interaction $J_2$. 
It is noteworthy that this model has recently been used to analyze the neutron measurement in the AC Au$_{72}$Al$_{14}$Tb$_{14}$, which results in $J_2/J_1\approx 2.3$ and $\theta=86^{\circ}$ that successfully reproduces not only the measured magnetic structure but also the temperature dependence of the magnetic susceptibility as well as the magnetic-field dependence of the magnetization~\cite{Sato2019}. 
Hence we adopt this model as the effective model, which is expected to be relevant generally to the Tb-based ACs and QCs. 
So far, in various Tb-based ACs, negative Weiss temperature was observed, which indicates that the antiferromagnetic (AF) interaction is dominant between the magnetic moments at the Tb sites~\cite{Suzuki}. 
Hence we focus on the AF interaction $J_1<0$ and $J_2<0$ and the case of the FM interaction will be reported elsewhere~\cite{W2021}. 

{
It is noted that the Tb-based AC and QC are metallic compounds. Hence, the RKKY interaction via the conduction electrons is considered to contribute to the exchange interaction between the 4f magnetic moments at Tb sites. Furthermore, in the AC, the effect of the Fermi surface such as the nesting can also contribute to trigger the magnetic order. To describe the essential feature of the resultant magnetism in the itinerant electron system, the minimal model is introduced as Eq.~(\ref{eq:H}). Namely, due to the strong magnetic anisotropy arising from the CEF, the Ising model restricted to the anisotropy direction at each Tb site, which is regarded as the strong anisotropy limit in the anisotropic Heisenberg model, is considered to be the effective model as the first step of analysis. 
}

\begin{figure}[h]
\includegraphics[width=8.9cm]{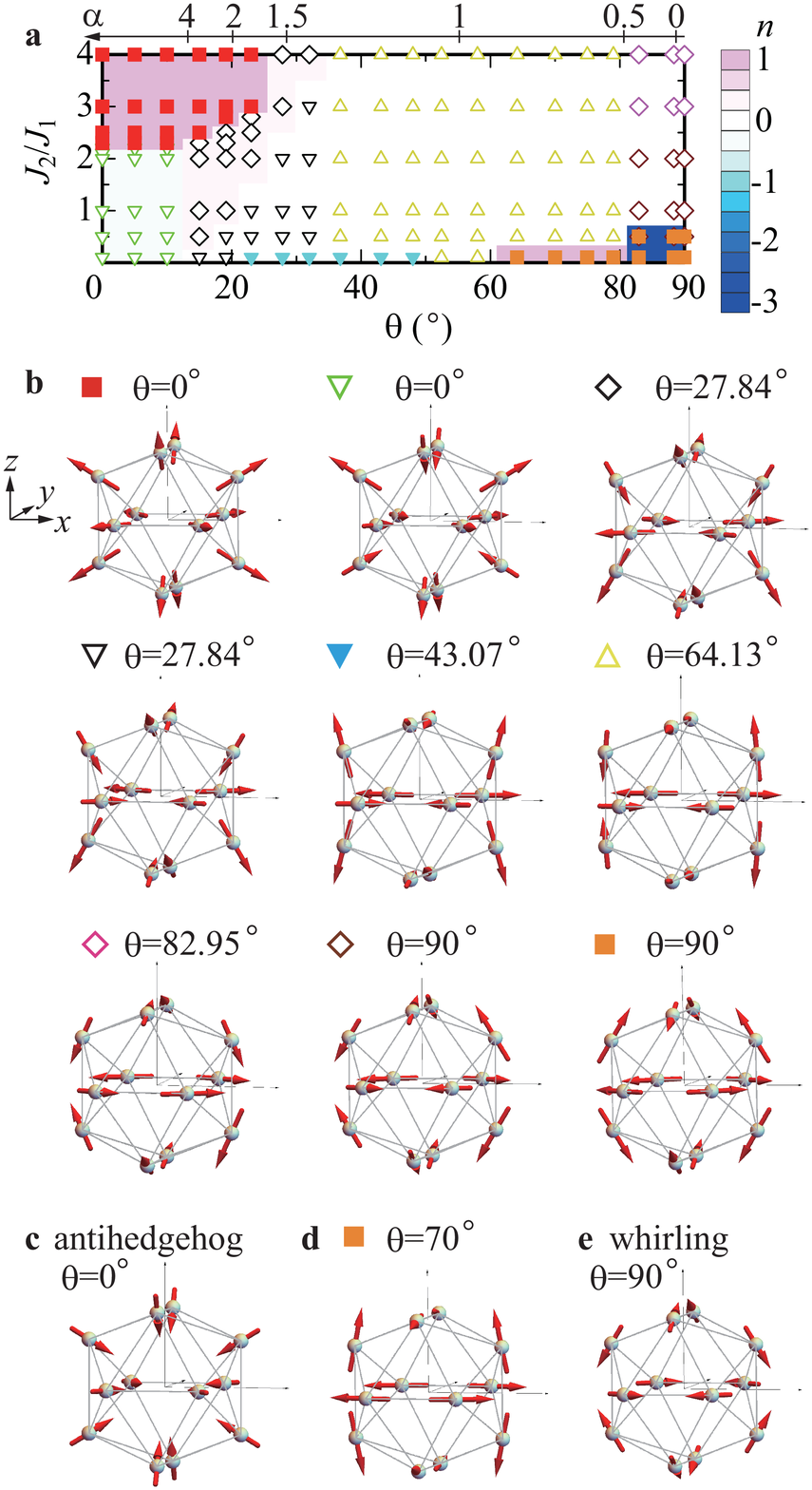}%
\caption{(color online) 
{\bf Ground-state phase diagram and magnetic textures in icosahedron.}  
{\bf a}, The ground-state phase diagram for $J_2/J_1$~$(J_1<0)$ and $\theta$ as well as $\alpha$ in the icosahedron. 
Contour plot of the topological charge $n$ is also shown. 
{\bf b}, Magnetic texture shown by each symbol in {\bf a}. 
{\bf c}, Antihedgehog state for $\theta=0^{\circ}$. 
{\bf d}, Antiwhirling state for $\theta=70^{\circ}$. 
{\bf e}, Whirling state for $\theta=90^{\circ}$. 
(This figure is created by using Adobe Illustrator CS5 Version 15.1.0.)
}
\label{fig:PD}
\end{figure}

\noindent
{\bf Ground-state phase diagram.} 
The ground-state phase diagram for $J_2/J_1$ and the angle $\theta$ of the magnetic anisotropy in the IC is shown in Fig.~\ref{fig:PD}a. Here we also show the $\alpha$ axis as the holizontal axis. It is noted that depending on the compositions of Au and SM in the Au-SM-Tb ACs and QCs, the $\alpha$ dependence of $\theta$ can change slightly from that shown in Fig.~\ref{fig:CEF}c. Hence, the $J_2/J_1$--$\theta$ phase diagram with the wide range of $\theta$ is generally relevant to the Tb-based ACs and QCs. 

\noindent
{\bf Magnetic textures on icosahedron.} 
We find that unique magnetic states appear depending on $J_2/J_1$ and $\theta$, as a result of the CEF-anisotropy-protected IC. 
Each symbol in Fig.~\ref{fig:PD}a represents the magnetic texture shown in Fig.~\ref{fig:PD}b. 
Note that the magnetic texture where all the moments are inverted also has the same ground-state energy for each symbol, i.e., energetically degenerate. 
The red-square symbols denote the hedgehog state where all the magnetic moments at 12 Tb sites on the IC are directed outward (see $\theta=0^{\circ}$ in Fig.~\ref{fig:PD}b). 
It is remarkable that the hedgehog state is stabilized in the rather wide-$\theta$ region (Supplementary information). 
The antihedgehog state where all the moments are inverted from the hedgehog state is shown in Fig.~\ref{fig:PD}c. 

\noindent
{\bf Topological charge.} 
To reveal the topological character underlying in the magnetic textures, we define the topological charge $n$ per an IC by calculating the solid angle subtended by three neighboring moments ${\bm J}_i$, ${\bm J}_j$, and ${\bm J}_k$~\cite{Eriksson} located at the vertices of each triangle surface of the IC. 
By calculating $n$ for each magnetic textures on the IC, we find that the topological charge of $n=1$ is realized in the hedgehog state and $n=-1$ is realized in the antihedgehog state. 
This implies the solid angle subtended by 12 Tb moments on the IC covers the whole sphere and 
the hedgehog and and antihedgehog play the roles of source or sink of emergent field, which are transcribed as the emergent monopole and antimonopole with a ``charge" $n=1$ and $n=-1$ respectively. 

\noindent
{\bf Whirling-moments state with unusually large topological charge.}
We find that the finite topological charge also appears in the region denoted by the symbols of orange squares in Fig.~\ref{fig:PD}a. 
The magnetic texture (see orange square $\theta=90^{\circ}$ in Fig.~\ref{fig:PD}b) is characterized by unusually large magnitude of topological charge $n=-3$. We call this state the antiwhirling-moments state since the magnetic moments are whirling when it is seen from the (111) direction (Supplementary information). 
Interestingly, we find the topological transition to $n=1$ occurs around $\theta=79^{\circ}$ (see Fig.~\ref{fig:PD}d). 
In Fig.~\ref{fig:PD}a, the contour plot of the topological charge $n$ is also shown. 
We see that finite integer $n$ appears in the ${\bm J}_{\rm tot}={\bf 0}$ state which is denoted by the square symbols in Fig.~\ref{fig:PD}a. 
Here, the total magnetic moment on the IC is given by ${\bm J}_{\rm tot}=\sum_{i=1}^{12}{\bm J}_i$. 
There exists a finite magnetic moment per an IC ${\bm J}_{\rm tot}\ne{\bf 0}$ for the magnetic textures denoted by not-square symbols in Figs.~\ref{fig:PD}a and \ref{fig:PD}b. 

\begin{figure}[h]
\includegraphics[width=8.9cm]{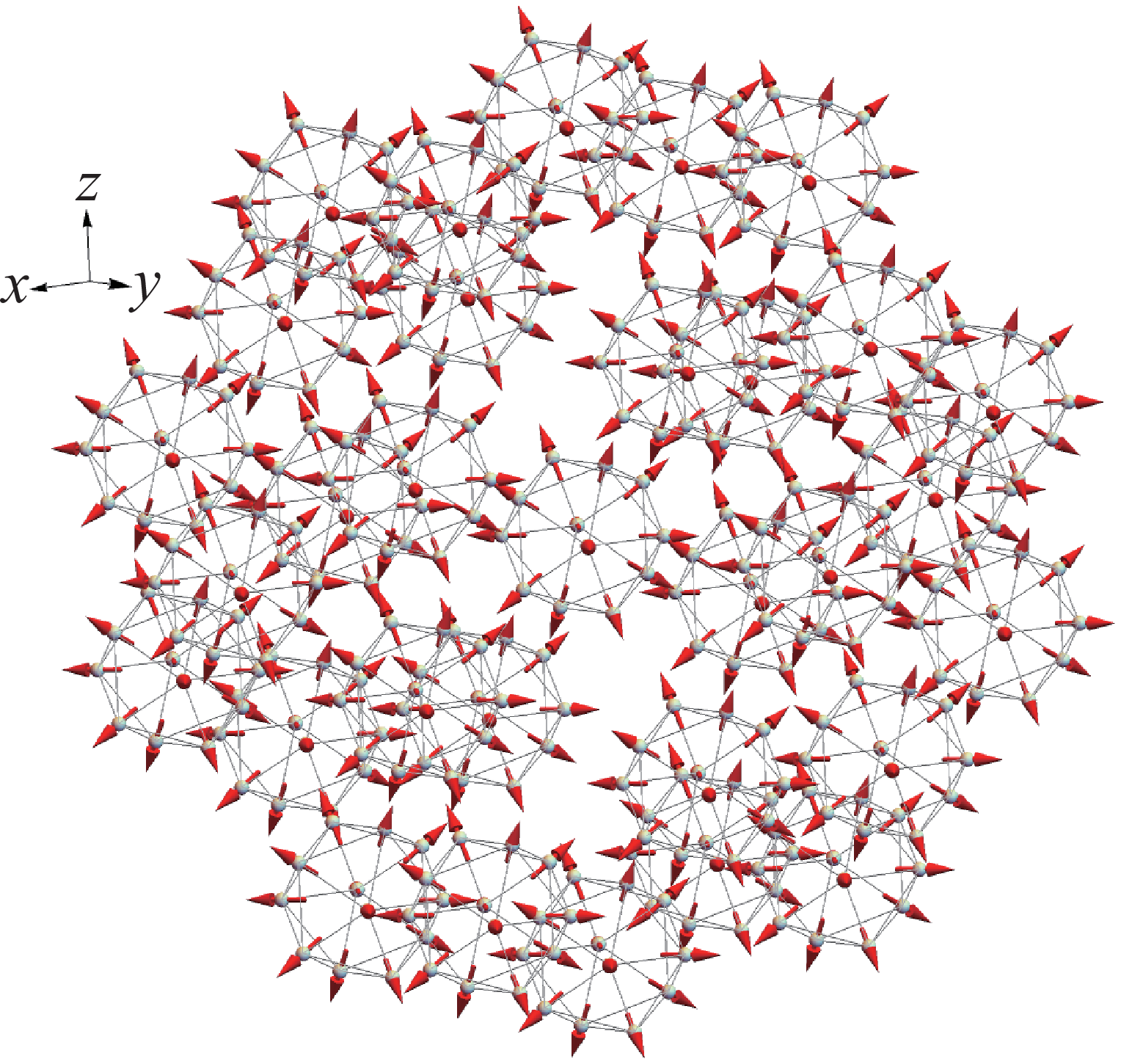}%
\caption{(color online) 
{\bf Long-range order of hedgehog in quasicrystal.} 
The long-range order of the hedgehog for $\theta=0^{\circ}$ in the QC. 
Tb-12 clusters at the origin and at the vertices of the icodidodecahedron are shown. 
(This figure is created by using Adobe Illustrator CS5 Version 15.1.0.)
}
\label{fig:HHQC}
\end{figure}

\noindent
{\bf Long-range order of hedgehog in quasicrystal.} 
To explore the magnetism in the QC, we consider the Cd$_{5.7}$Yb-type QC where 
the Tb-12 cluster i.e., IC is located at the origin and the surrounding vertices of the icosidodecahedron and they are repeatedly arranged outward in the self-similar manner~\cite{Takakura}. 
We have applied the model~(\ref{eq:H}) to the Cd$_{5.7}$Yb-type QC with the AFM N.N. interaction $J_1<0$ and N.N.N. interaction $J_2<0$ not only for the intra IC but also for the inter ICs located on the line connecting neighboring vertices of the $\tau^3$-times enlarged icosidodecahedron from that in Fig.~\ref{fig:atoms}d. 
Here $\tau=(1+\sqrt{5})/2$ is the golden ratio. 
%
{
Then we evaluate the inner product $\langle\hat{\bm J}_i\cdot\hat{\bm J}_j\rangle$ for the N.N. and N.N.N. pairs of the Tb sites in the intra IC and inter ICs. For the hedgehog state with $\theta=0^{\circ}$, we confirmed that $\langle\hat{\bm J}_i\cdot\hat{\bm J}_j\rangle<0$ is realized for all the $\langle i,j\rangle$ pairs with respect to the inter ICs in addition to the intra IC. This indicates that the hedgehog is stabilized in the arrangement of the 30 ICs located at each vertex of the icosidodecahedron. Indeed we confirmed that the hedgehog has the lowest ground-state energy. }
We show the magnetic structure in the real space for $\theta=0^{\circ}$ in Fig.~\ref{fig:HHQC}. The hedgehog is arranged ferromagnetically i.e., uniformly distributed in IC to IC, where the total magnetic moment is zero ${\bm J}_{\rm tot}={\bf 0}$ in each IC.

%
{
This result is intuitively understood from the magnetic moment directions shown in Fig.~\ref{fig:HHQC}. Namely, each magnetic moment at the Tb site on the IC are directed outward. Hence, the magnetic moments between the neighboring ICs are directed antiferromagnetically each other at least for the N.N. and N.N.N. Tb sites when the hedgehog is uniformly distributed. Then, if the N.N. and N.N.N. interactions are AFM in the model (\ref{eq:H}), i.e., $J_1<0$ and $J_2<0$, the hedgehog in the IC is able to be arranged uniformly as shown in Fig.4. For the outer ICs than the icosidodecahedron, the magnetic moments pointing to outward direction from the IC earn the energy gain for the neighboring IC with the same hedgehog when the N.N. and N.N.N. interactions are AFM. Then, the uniform arrangement of the hedgehog state is realized in each IC in the self-similar manner, giving rise to the magnetic long-range order in the QC. }

%
This implies that the long-range order of the hedgehog is realized in the ground state. By calculating $\langle\hat{\bm J}_i\cdot\hat{\bm J}_j\rangle$ for N.N. and N.N.N. pairs of the Tb sites for the inter ICs in 30 ICs located at vertices of the icosidodecahedron, we confirmed that uniform distribution of the hedgehog is stabilized as the ground state at least for the region of $J_2/J_1$ and $\theta$ symbolized by red squares in Fig.~\ref{fig:PD}a.
This is, to our best knowledge, the first discovery of the 
%
{possible} 
long-range order of topological magnetic textures in the QC. 
It is noted that the antihedgehog shown in Fig.~\ref{fig:PD}e also exhibits the uniform 
%
{distribution}.
The hedgehog QC and antihedgehog QC are energetically degenerate within the model (\ref{eq:H}). 

\noindent
{\bf Discussion.} 
Our finding of the hedgehog long-range order in the QC demonstrates that topological textures can be stabilized in the QC. 
As for the antiwhirling (whirling) state [see orange square for $\theta=90^{\circ}$ in Fig.~\ref{fig:PD}b (Fig.~\ref{fig:PD}e)], the long-range order can be realized uniformly when the AFM N.N.N. interaction is dominant for the inter ICs in the QC (Supplementary information).
In this case, the antimonopole (monopole) with unusually large topological charge $|n|=3$ orders ferromagnetically. In the hedgehog and whiling ordered states, the non-trivial phenomena such as the topological Hall effect are expected to emerge. 
Our model (\ref{eq:H}) is generically applicable to the rare-earth based QCs and ACs with magnetic anisotropy, which is useful to clarify the magnetic structures as well as topological properties.


\newpage



{\bf Methods}

{\bf Crystalline electric field}

Recently, the CEF Hamiltonian in the rare-earth based QCs and ACs have been formulated by the operator-equivalent methods~\cite{WM2021} as
\begin{eqnarray}
H_{\rm CEF}=\sum_{\ell=2,4,6}\left[
B_{\ell}^{0}(c)O_{\ell}^{0}(c)+\sum_{\eta=c,s}\sum_{m=1}^{\ell}B_{\ell}^{m}(\eta)O_{\ell}^{m}(\eta)
\right],
\label{eq:HCEF}
\end{eqnarray}
where $O_{\ell}^{m}(\eta)$ is the Stevens operator~\cite{Stevens} and $B_{\ell}^{m}$ is the coefficient. 
Since $O_{\ell}^{m}(\eta)$ is expressed as the operators of the total angular momentum $J$, $H_{\rm CEF}$ can be obtained on the basis of $J$ and the $z$ component $J_z$, i.e., $|J, J_z\rangle$ for $J_z=J, J-1, \cdots, -J+1$, and $-J$. Then, by diagonalizing the $2J+1\times 2J+1$ matrix $H_{\rm CEF}$, we obtain the CEF energies $E_n$ and the eigenstate $|\psi_n\rangle$. For the Tb$^{3+}$ ion, $J=6$ is the ground multiplet according to the Hund's rule. Then, $H_{\rm CEF}$ is diagonalized by setting $J=6$ as described in the main text (see also Supplementary information). 

{\bf Principal axis of the magnetization}

The largest magnetic-moment direction in the CEF ground state can be identified by diagonalizing the $3\times 3$ matrix $M_{\xi\delta}=\langle\psi_0|\hat{J}_{\xi}\hat{J}_{\delta}|\psi_0\rangle$ for $\xi, \delta=x, y$, and $z$. 
Here, $\hat{J}_{\xi}$ is the operator of the total angular momentum. 
The normalized eigenvector of the largest eigenvalue of $M$, ${\bm J}=(J_x, J_y, J_z)$, gives the direction, i.e., the principal axis of the magnetic moment.

{\bf Minimal model}

The ground state of the model (\ref{eq:H}) in the main text is obtained numerically among $2^{12}$ spin configurations in the IC. 
%
Namely, the expectation value of the Hamiltonian (1), i.e., the energy $E=\langle \sigma_1,\sigma_2,\cdots,\sigma_{12}|H| \sigma_1,\sigma_2,\cdots,\sigma_{12}\rangle$ are calculated for $\sigma_i=\uparrow, \downarrow$ where the quantization axis at each $i$-th site on the IC (Fig.~\ref{fig:atoms}c) is directed to the anisotropy direction $\theta$ shown in the inset of Fig.~\ref{fig:CEF}. Then, by searching the lowest energy numerically, we obtain the ground state. By performing this calculation for various $J_2/J_1$ and $\theta$, we have constructed the ground-state phase diagram for an IC in Fig.~\ref{fig:PD}.

%
Next, we apply the model (\ref{eq:H}) to the Cd$_{5.7}$Yb-type QC, where the Tb-12 cluster, i.e., IC, is located at the origin surrounded by the $\tau^3$ times enlarged icosidodecahedron from that in Fig.1d, whose vertices ICs are located at and they are repeatedly arranged outward in the self-similar manner.
In the icosidodecahedron, there exist 30 ICs. For the neighboring pairs of ICs, whose number is 60, we set the N.N. interaction $J_1$ and N.N.N. interaction $J_2$ between the magnetic moments at the N.N. and N.N.N. Tb sites respectively. 

%
Then, we evaluate the inner product $\langle\hat{\bm J}_i\cdot\hat{\bm J}_j\rangle$ for four sets of $\langle i,j\rangle$ pair as the N.N. and N.N.N. pairs per each bond among the 60 bonds. Namely, the inner product for the $4\times 60$ pairs are calculated with respect to the inter ICs. Combining this with the calculation for the intra IC mentioned above, we search the lowest-energy state of the model (\ref{eq:H}) on the 30 ICs located at vertices of the icosidodecahedron with totally 360 Tb sites numerically for various $J_2/J_1$ and $\theta$ and obtain the ground state. 

{\bf Topological charge on the IC}

%
Recently, the scalar chirality and the topological charge have been defined for four spins located at each vertex of the tetrahedron~\cite{Aoyama}. It has been shown by this definition that the hedgehog state realized in the tetrahedron is characterized by the topological charge of one. Extending these definitions to the IC, here we consider the scalar chirality and the topological charge in the IC as follows.

We define the scalar chirality of the IC 
with its center-of-mass position ${\bm R}$ by $\chi({\bm R})=\sum_{i,j,k\in{\rm IC}}\chi_{i,j,k}$ with $\chi_{i,j,k}\equiv {\bm J}_i\cdot({\bm J}_j\times{\bm J}_k)$ where the order of $i$, $j$, and $k$ is defined in the anticlockwise direction with respect to the normal vector of the triangle formed by the $i$, $j$, and $k$th sites, $\hat{n}_{i,j,k}$, pointing outward from ${\bm R}$. 
In the same way, we define the solid angle subtended by the twelve moments on the IC as 
$\Omega({\bm R})=\sum_{i,j,k\in{\rm IC}}\Omega_{ijk}$. 
Here, $\Omega_{ijk}$ is the solid angle for three magnetic moments ${\bm J}_i$, ${\bm J}_j$, and ${\bm J}_k$, 
which is given by $\Omega_{ijk}=2\tan^{-1}[\chi_{ijk}/(1+{\bm J}_i\cdot{\bm J}_j+{\bm J}_j\cdot{\bm J}_k+{\bm J}_k\cdot{\bm J}_i)]$~\cite{Eriksson}. 
The topological charge $n$ is defined as $n\equiv\Omega({\bm R})/(4\pi)$ per an IC. 

{\bf Data availability}

All the data supporting the findings are available from the corresponding author upon reasonable request.


{\bf Acknowledgments}

This work was supported by JSPS KAKENHI Grant Numbers JP18K03542 and JP19H00648.

{\bf Author contributions}

S.W. conceived the study and led the project. Theoretical calculation was performed by S.W. The manuscript was written by S.W.

{\bf Competing interests}

The author declares no competing interests.

{\bf Additional information}

{\bf Supplementary information} is available for this paper. 

{\bf Correspondence and requests for materials} should be addressed to S.W.

\end{document}